\documentclass[twocolumn,showpacs,preprintnumbers,amsmath,amssymb,epsfig,widetext]{revtex4}
\newcommand{\ead}{\email}
\usepackage{epsfig}

\newcommand{\bn}{\hat {\bf n}}
\newcommand{\bl}{{\bf l}}

\begin{document}

\title[Mass Estimators from CMB]{Cluster Mass Estimators from CMB Temperature
and Polarization Lensing}

\author{Wayne Hu,$^{1}$ Simon De Deo,$^{1}$ Chris Vale$^{2}$}

\address{$^{1}$Kavli Institute for Cosmological Physics, Department of Astronomy and Astrophysics,
and Enrico Fermi Institute,University of Chicago, Chicago IL 60637}
\address{$^{2}$Particle Astrophysics Center, Fermilab, P.O. Box 500, Batavia, IL 60510} 
\ead{whu@background.uchicago.edu}
\begin{abstract}
Upcoming Sunyaev-Zel'dovich surveys are expected to return $\sim 10^4$ intermediate
mass clusters at high redshift.  Their average masses must be known to same accuracy as
desired for the dark energy properties.  Internal to the surveys, the 
CMB potentially provides a source for lensing mass measurements whose distance is precisely 
known and behind all clusters.   
We develop statistical mass estimators 
from 6 quadratic combinations of CMB temperature and polarization
fields that can simultaneously recover large-scale structure and cluster mass profiles.
The performance of these estimators on idealized
NFW clusters suggests that surveys with a $\sim 1'$ beam and 
$10\mu$K$'$ noise in uncontaminated temperature maps
can make a $\sim 10\sigma$ detection, or equivalently a $\sim 10\%$ mass measurement
 for each $10^3$ set of clusters. With internal or external acoustic scale 
$E$-polarization measurements, the $ET$ cross  
correlation estimator can provide a stringent test for contaminants on a first 
detection at $\sim 1/3$ the significance. 
For surveys that reach below $3\mu$K$'$, the $EB$ cross correlation estimator should provide
the most precise measurements and potentially the strongest control over contaminants.
\end{abstract}

\maketitle

\section{Introduction}

Upcoming surveys for clusters utilizing the Sunyaev-Zel'dovich (SZ) effect in the
Cosmic Microwave Background (CMB) as a detection technique hold the promise to measure 
the properties of the dark energy to high precision with $\sim 10^4$ clusters at 
redshifts out to $z\sim 1$.    The mean mass and variance of the sample needs to be 
known to an accuracy comparable to that desired for the dark energy equation of state 
to not be a limiting factor.   Since the SZ effect is sensitive to the temperature weighted baryon 
content of the cluster, it does not directly probe the total cluster mass in a model 
independent way.  

Fortunately, the same survey which identifies the clusters in the SZ effect can potentially
also constrain their masses through gravitational lensing of the CMB.  Utilizing the
CMB as a source is also appealing in that its distance is both well determined and
sufficiently far to probe even the highest redshift clusters in the sample.

Studies of gravitational lensing of the CMB by clusters 
\cite{SelZal00,NagKraKos03,Dodelson04,HolKos04,ValAmbWhi04,MatBarMenMos05,LewKin06} 
typically focus on making detailed measurements of individual high mass clusters or on 
statistical reconstructions of average cluster properties.  The CMB 
suffers in the former case when compared with galaxy weak lensing because the background 
image, in this case the primary CMB, is a Gaussian random field whose properties must 
be separated from the cluster and its emission.     

This disadvantage will be substantially reduced in the latter case where the goal 
is to measure statistical properties rather than unique objects, provided that large 
samples of clusters are available for the analysis.  Surveys such as 
SPT and ACT should provide samples of $\sim 10^{4}$ intermediate mass clusters at high 
redshift, so that a precise and unbiased statistical measurement may become a 
realistic possibility in the near future.  The CMB even possesses some advantages over 
galaxy lensing, since the source statistical properties and redshift are extremely well 
determined, and because lensing of the temperature and polarization provide strong 
consistency checks against possible contamination by cluster emission. 

Techniques which use CMB lensing to reconstruct the lensing convergence, $\kappa$, have 
been available for some time \cite{Hu01b,HuOka01,HirSel02}, and although these were 
initially intended as probes of large scale structure, they can in principle recover the 
projected mass regardless of its source.  This certainly includes reconstructions of large 
galaxy clusters and associated statistical quantities, such as the cluster-convergence 
correlation function.  On cluster length scales, this is essentially an average density profile 
and mass measurement (e.g. \cite{GuzSel01,Johetal05}).  We will focus on estimators that
are quadratic in the observed CMB fields \cite{Hu01b,HuOka01}, since these can be 
implemented using fast algorithms, and can eliminate contamination by isolating pairs of modes.

However, Amblard {\it et al.} \cite{AmbValWhi04} used simulations to show that the minimum 
variance quadratic estimator built out of the temperature field is biased if the lensing 
field is non-Gaussian at the level expected for real structure.  Subsequently Maturi {\it et al.}~\cite{MatBarMenMos05} helped to explain this fact by deomonstrating that the reconstructed 
density field for this estimator is biased low in regions around large clusters.

In this 
paper, we analyze the origin of the bias and show that it can be nearly eliminated 
using the increasingly well measured properties of the CMB, and that the modifications we 
introduce produce only a very modest degradation in the statistical reconstruction of cluster 
properties.  Since the temperature field estimators readily 
generalize to polarization, these can be used as consistency checks on each other or indeed 
on any other cluster mass estimates.

The outline of the paper is as follows. In \S \ref{sec:quadratic} we review the general
construction of quadratic estimators and the underlying approximations upon which
they are based.  We determine the origin of the biased estimates and show that they
are associated with the miss-estimation of source gradients in the moderate to
strong lensing regime.   The bias can be eliminated at negligible cost to the signal
to noise by imposing a strong filter against this and other contaminants.  In \S \ref{sec:examples},
we consider the idealized performance of the temperature and polarization estimators.

For illustrative purposes, we use throughout a flat $\Lambda$CDM cosmology with
matter density $\Omega_m=0.24$, baryon density $\Omega_b h^2=0.0223$,
Hubble constant $h=0.73$, scalar tilt $n_s=0.958$, amplitude $\sigma_8=0.76$.

\section{Quadratic Estimators}
\label{sec:quadratic}

Lensing is a surface brightness conserving remapping of the intrinsic temperature
and polarization fields from recombination.    
Given an unlensed temperature field $\tilde T(\bn)$ and Stokes $\tilde Q(\bn)$ and 
$\tilde U(\bn)$ 
fields, where
$\bn$ denotes the angular position on the sky, the lensed fields are given by
\begin{eqnarray}
T(\bn) &=&  \tilde T(\bn + \nabla \phi) \,, \nonumber\\
Q(\bn) &=&  \tilde Q(\bn + \nabla \phi) \,, \nonumber\\
U(\bn) &=&  \tilde U(\bn + \nabla \phi) \,.
\label{eqn:mapping}
\end{eqnarray}
Here $\phi$ is the deflection potential, $\nabla \phi$ is the
deflection angle, and they are related to the convergence as 
\begin{equation}
\nabla^2 \phi = -2 \kappa \,.
\end{equation}
All derivatives throughout are angular derivatives on the sky.  We will furthermore
employ the flat sky
approximation in this paper but all expressions readily generalize to the
full sky with the replacement of ordinary derivatives with covariant derivatives and Fourier
modes with spherical harmonics
\cite{Hu01,ChaCho02,OkaHu03}.

The estimators of  $\phi$ or equivalently $\kappa$ 
introduced in \cite{Hu01b,HuOka01} rely on two related but independent linearization approximations: 
the gradient approximation
and linearization in the convergence.  Only the latter, which is equivalent to considering lensing
as a small correction to the source field, requires modification
for use in cluster reconstruction. 

To see how these approximations enter, let us first consider the case of the temperature 
field.  The analysis readily generalizes to polarization as we shall see.

Fundamentally the temperature estimator is built out of a lensing induced
correlation between the temperature field and its gradient \cite{Zal00}.  This correlation arises
from the gradient approximation, valid when the deflections are small compared with the
structure in the unlensed field.   This approximation of course does not hold for
 all Fourier modes in the CMB fields and lens  (see e.g. \cite{ChaLew05}).  It need only
hold the modes that are correlated by the reconstruction.
The lensed field $T_L(\bn)$
can be prefiltered in Fourier space to isolate modes for which the gradient approximation is
valid   
\begin{eqnarray}
L^T(\bn) &\equiv & T_{L}({\bn}) \,, \nonumber\\
T_L(\bn) &= & \int {d^2 l \over (2\pi)^2} e^{i \bl \cdot \bn} W_l^T T_\bl \,,
\end{eqnarray}
where $W_l^T$ is the Fourier filter, the subscript ``L" denotes the lensing-filtered field and
$T_\bl$ is the Fourier representation of the full temperature field.

The lensed temperature field $T_L$ can be approximated by a Taylor expansion
\begin{equation}
T_L  \approx \tilde T_L + (\nabla \tilde T \cdot \nabla \phi)_L \,,
\label{eqn:gradapprox}
\end{equation}
which we will call the gradient approximation.
For the case of cluster
lensing, the typical deflections are $< 1'$ compared with structure in the
unlensed CMB with coherence of $\sim 10'$.   Eq.~(\ref{eqn:gradapprox}) is therefore
an excellent approximation for the full field
even in the strong lensing regime.  In this case the minimum
variance filter can be used \cite{Hu01b}
\begin{equation} 
W_l^{T} = (C_l^{TT}+ N_l^{TT})^{-1} \,.
\label{eqn:lensfilter}
\end{equation}
Here $C_l^{TT}$ is the lensed CMB power spectrum and $N_l^{TT}$ is the noise power
spectrum.  More generally $W_l^T$ can be chosen to suppress modes which violate the gradient approximation
or which are contaminated by foregrounds.

Given that the gradient approximation induces a correlation with the unlensed temperature gradient,
one forms a quadratic estimator by multiplying $T_L$ by a filtered gradient of the {\it lensed} temperature
field
\begin{eqnarray}
{\bf G}^T(\bn) &\equiv& \nabla T_{G}(\bn)\,, \nonumber\\
T_G(\bn) &= & \int {d^2 l \over (2\pi)^2} e^{i \bl \cdot \bn} W_l^{TT} T_\bl\,,
\end{eqnarray}
where $W_l^{TT}$ is another Fourier space filter. The  filter \cite{Hu01b}
\begin{equation}
W_l^{TT} = {\tilde C_l^{TT}} (C_l^{TT}+ N_l^{TT})^{-1}
\label{eqn:gradfilterorig}
\end{equation}
yields a minimum variance reconstruction under the  fully  linearized approximation.
Here $\tilde C_l^{TT}$ is the unlensed CMB power spectrum and Eq.~(\ref{eqn:gradfilterorig})
is essentially
a Wiener filter for the unlensed gradients (c.f. \cite{MatBarMenMos05} for the
more problematic Wiener filter for the total gradient).

This filter has to be modified in the presence of rare non-Gaussian structure where lensing effects
are moderate to strong.   Expanding the product in the gradient approximation, we obtain
\begin{eqnarray}
{\bf G}^T L^T  &\approx & (\nabla\tilde T_G )\tilde T_L +   (\nabla \tilde T_G) (\nabla \tilde T \cdot \nabla \phi)_L \nonumber\\
&& +  [ (\nabla \phi) \cdot \nabla) \nabla\tilde T +
(\nabla \tilde T \cdot \nabla) \nabla\phi)]_G \tilde T_L \\
&&+ [ (\nabla \phi) \cdot \nabla) \nabla\tilde T +
(\nabla \tilde T \cdot \nabla) \nabla\phi)]_G (\nabla \tilde T \cdot \nabla \phi)_L \,. \nonumber
\end{eqnarray}
The second approximation is that the quadratic term in $\phi$  is negligible. 
If so, averaging over realizations of the unlensed temperature field yields a quantity
proportional to $\nabla \phi$ with a proportionality coefficient related to the well-determined
two-point function or power spectrum of the unlensed CMB.   This quantity is therefore a quadratic
estimator of the deflection angles $\nabla\phi$ or  $\kappa$
in the linear approximation.

The quadratic term in $\phi$ comes from the change in the
temperature gradients due to lensing.  Qualitatively, its omission is equivalent to considering
lensing as a small perturbation to the unlensed image.   Quantitatively it involves the assumption
that the deflection angles are small compared with structures in the lens.
This approximation is violated around a cluster and the Wiener filter, which is based
on the variance of typical regions, does not sufficiently suppress this region in the absence
of detector noise.  

To see the effect of the quadratic term, consider the lensed temperature field to be filtered
for small scale fluctuations as in Eq.~(\ref{eqn:lensfilter}) with $N_l^{TT}=0$ and likewise take the
unlensed temperature gradients $\nabla \tilde T_G \approx$ const.  In that case, the estimator
becomes 
\begin{eqnarray}
{\bf G}^T L^T   \approx
\left[ \nabla \tilde T +  (\nabla \tilde T \cdot \nabla) \nabla\phi)\right]_G   (\nabla \tilde T \cdot \nabla \phi)_L \,.
\end{eqnarray}
The quadratic term carries a coherent contribution whose strength depends on $\kappa$
since $\nabla^2\phi = -2\kappa$.   As $\kappa$
becomes ${\cal O}(1)$ in the interior of the cluster, the estimation will be biased low.
In fact for $\kappa > 1$ the observed gradient reverses and a second, flipped image of
the intrinsic gradient appears at the center of an azimuthally symmetric cluster.

Another way to see why the reconstruction is biased low in the single image regime
is that the cluster magnifies
the background image and decreases the observed temperature gradient behind it.
While this bias is mitigated by the Wiener filtering of Eq.~(\ref{eqn:gradfilterorig}),  the filter
cannot remove it entirely.    Furthermore, a truly optimal filter would require knowledge
of the cluster mass to be estimated.

The bias arises because of the overlap in scales between the unlensed gradient field
and the lensed temperature field.  
Maturi et al. \cite{MatBarMenMos05} suggested that it can be further mitigated by
a combination of more strongly high pass filtering the lensed temperature field in Eq.~(\ref{eqn:lensfilter})
 and utilizing
the direction of the large scale gradients.   The efficacy of this technique depends on both the assumed
signal and the instrument noise.

\begin{figure}[th]
\centerline{\psfig{file=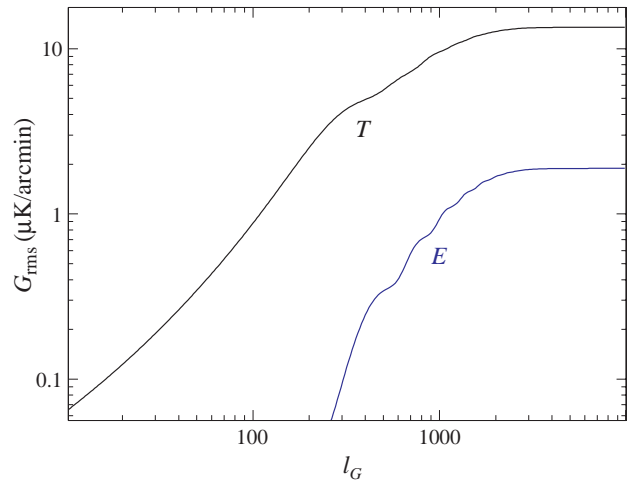, width=3.25in}}
  \caption{Gradient of the unlensed temperature $T$ and polarization $E$ fields as a function of
  the maximum $l=l_G$ of a top hat low pass filter.  For both fields, gradients are dominated by
  modes with $l<2000$.}
  \label{fig:grad}
\end{figure}

Instead let us exploit the fact that we have robust prior knowledge about the
unlensed CMB spectrum.   A handful of well determined cosmological parameters fixes
its shape out through the damping tail where the small scale gradients pick up most of their
contribution.   Fig.~\ref{fig:grad} shows the unlensed rms gradient as a function of a step function low
pass filter $W^{TT}_l$ out to $l=l_G$ 
\begin{eqnarray}
(G_{{\rm rms}}^T)^2 &\equiv& \langle \nabla \tilde T_G(\bn) \cdot \nabla \tilde 
T_G(\bn) \rangle \nonumber\\
                   &=& \int_1^{l_{G}} { l d l \over2\pi} l^2 \tilde C_l^{TT} \,.
                   % \nonumber\\
%G_{E{\rm rms}}^2(l_{G}) &\equiv& \langle \nabla E_G(\bn) \cdot \nabla E_G(\bn) \rangle \nonumber\\
 %                  &=& \int_1^{l_{G}} { l d l \over2\pi} l^2 C_l^{EE} \nonumber
\end{eqnarray}                
Almost all  of the gradient comes from $l \le 2000$ due
to diffusion damping of the acoustic peaks.  On these scales, the lensing correction to the gradient
is tiny for a typical cluster.  This suggests that if we impose a sharp filter
on the gradient to exclude higher multipoles, we lose very little signal and gain a clean
separation between the lensed and unlensed structure.    Such a filter is also
desirable for large scale reconstruction to eliminate pairs of modes that violate the
gradient approximation.  Furthermore, it also reduces contamination from foregrounds and systematics
that might appear as signal without a clean separation of scales between the two fields
of the quadratic estimator.

Now let us make these considerations concrete and generalize them to the full set of temperature and polarization fields
$X,Y \in T,E,B$.    $E(\bn)$ and $B(\bn)$ are real valued fields that are related to the Stokes parameters as
\begin{eqnarray}
Q(\bn)+i U(\bn) = - \int {d^2 l \over (2\pi)^2} [E_\bl+ i B_\bl] e^{2i \varphi_{\bl}} e^{i \bl \cdot \bn}\,,
\end{eqnarray}
where $\varphi_{\bl}$ is the angle between $\bl$ and the axis which defines Stokes $Q$.
We shall assume that $B$ is absent in the unlensed CMB.

The angular space convergence estimators 
\begin{equation}
\hat \kappa^{XY}(\bn) = \int {d^2 l \over (2\pi)^2} e^{i\bl \cdot \bn} \hat \kappa^{XY}_{\bl}\,,
\end{equation}
are constructed from Fourier space estimators that are 
quadratic in the observed fields
\begin{equation}
{\hat \kappa^{XY}_{\bl} 
\over A_l^{XY}}
 =- \int {d^2\bn } e^{-i \bn \cdot \bl} {\rm Re} \left\{ \nabla
\cdot [  {\bf G}^{XY} (\bn) L^{Y*} (\bn)]  \right\}\,.
\label{eqn:fourierestimator}
\end{equation}
The gradient field ${\bf G}^{XY}$ is built out of the lensed $X$ field as either a spin-0 or spin-2 field
depending on the spin of $Y$
\begin{equation}
%G^{TT}(\bn) & =&    \int {d^2 l  \over (2\pi)^2} e^{i \bn \cdot \bl} W^{TT}_l{T(\bn) \over C_l^{TT}+N_l^{TT} } \nonumber\\
{\bf G}^{XY}_\bl =   i \bl \, W^{XY}_l {X_\bl}  \times \Big\{
\begin{array}{lr}
1\,, & Y=T\,, \\
e^{2i\varphi_{\bl}} \,, & Y=E,B\,.\\
\end{array}
\end{equation}
$L^Y$ represents the lensed fields
\begin{equation}
L^{Y}_\bl =   W^{Y}_l {Y_\bl}  \times \Big\{
\begin{array}{lr}
1\,, & Y=T\,, \\
e^{2i\varphi_{\bl}} \,, & Y=E,B\,.
\end{array}
\end{equation}
For the polarization fields, they are 
the Stokes field $Q + i U$ filtered for the $E$ and $B$ components.
The 2 gradient fields and 3 lensed fields produce 6 estimators from the temperature
and polarization fields for consistency checks.   In addition, replacement of the
divergence in Eq.~(\ref{eqn:fourierestimator})  with curl should leave estimators that are
consistent with noise.  

$A_l^{XY}$ is a normalization coefficient set to return unbiased estimators under
the fully linearized approximation.
For arbitrary filter functions, it is determined by noting that \cite{HuOka01}
\begin{equation}
\langle X_{\bl_1} Y_{\bl_2} \rangle = 2  f^{XY}(\bl_1,\bl_2)  \kappa_\bl\,,
\end{equation}
where  $\bl = \bl_1+\bl_2 \ne 0$ and
\begin{eqnarray}
f^{TT}(\bl_1,\bl_2) &=& \tilde C_{l_1}^{TT} (\bl \cdot \bl_1) + \tilde C_{l_2}^{TT} (\bl \cdot \bl_2) \,,\nonumber\\
f^{TE}(\bl_1,\bl_2) &=& \tilde C_{l_1}^{TE}\cos 2\varphi_{\bl_1\bl_2}
 (\bl \cdot \bl_1) + \tilde C_{l_2}^{TE} (\bl \cdot \bl_2)\,, \nonumber\\
f^{TB}(\bl_1,\bl_2) &=& \tilde C_{l_1}^{TE} \sin 2\varphi_{\bl_1\bl_2}  (\bl \cdot \bl_1)\,,\nonumber\\
f^{ET}(\bl_1,\bl_2) &=& f^{TE}(\bl_2,\bl_1)\,, \nonumber\\
f^{EE}(\bl_1,\bl_2) &=& [\tilde C_{l_1}^{EE}
 (\bl \cdot \bl_1) + \tilde C_{l_2}^{EE} (\bl \cdot \bl_2)] \cos 2\varphi_{\bl_1\bl_2} \,,\nonumber\\
f^{EB}(\bl_1,\bl_2) &=& \tilde C_{l_1}^{EE} \sin 2\varphi_{\bl_1\bl_2}  (\bl \cdot \bl_1)\,.
\end{eqnarray}
Here $\varphi_{\bl_1 \bl_2} \equiv \varphi_{\bl_1} - \varphi_{\bl_2}$.
The normalization is given in terms of these quantities and the filters as
\begin{eqnarray}
{ 1\over A_l^{XY} }=   {2 \over l^2} \int{ d^2 l_1 \over(2\pi)^2} ({\bl \cdot \bl_1})
 W_{l_1}^{XY} W_{l_2}^Y c^Y({\bf l}_1,{\bf l}_2) \nonumber
% \nonumber\\
% && \quad\times
  f^{XY}({\bf l}_1,{\bf l}_2)\,,
\end{eqnarray}
where
\begin{eqnarray}
c^T(\bl_1,\bl_2) &=& 1\,, \nonumber\\
c^E(\bl_1,\bl_2) &=& \cos 2\varphi_{\bl_1 \bl_2} \,,\nonumber\\
c^B(\bl_1,\bl_2) &= & \sin 2\varphi_{\bl_1 \bl_2} \,.
\end{eqnarray}

For the lensed field filter, we retain the choice of \cite{Hu01b,HuOka01}
\begin{eqnarray}
W^Y_l  =  ( C_l^{YY}+N_l^{YY})^{-1} \,.
\end{eqnarray}
As described above, the first modification is that
the gradient weights are set to zero above $l_G=2000$
\begin{equation}
W_l^{XY} = 0\,, \qquad (l > l_G) \,.
\end{equation}
Fig.~\ref{fig:grad} shows that this is also a good choice for the $E$ field.
For $l\le l_G$, we retain the Wiener filter
\begin{equation}
W_l^{XY} =
% \Big\{
%\begin{array}{lr}
 \tilde C_l^{XY} (C_l^{XX}+N_l^{XX})^{-1}\,,  
%&  
\quad (l\le l_G)\,,
%0\,, & l>l_G\,, \\
%\end{array}
\end{equation}
for $Y \in T,E$ and 
\begin{equation}
W_l^{XB} = 
%\Big\{
%\begin{array}{lr}
 \tilde C_l^{XE} (C_l^{XX}+N_l^{XX})^{-1}\,,  
 \quad (l\le l_G)\,,
 %&  l\le l_G\,,\\
%0\,, & l>l_G\,. \\
%\end{array}
\end{equation}
for $Y=B$. 

Secondly, we distinguish between $TE$ and $ET$ estimators.    With the gradient filter
employing only $l < l_G$, the scale symmetry between the two fields is broken.  It is
then advantageous to separate the estimators.   The $TE$ estimator uses $T$ for the gradient
field and $E$ for the lensed field.  If contamination from unpolarized cluster emission is strong
then this estimator can be used to eliminate its effects.  The $ET$ estimator uses $E$ for the
gradient field and $T$ for the lensed field.   This estimator places less demanding requirements on 
the experiments in that only the relatively large $E$ field of the acoustic peaks need be measured to high
signal to noise.   Indeed this measurement need not come from the same experiment that measures
the cluster signal.  Both estimators strongly reject contaminants from foregrounds and systematics.
Only the background CMB will have $T$ and $E$ correlated and anticorrelated in the specific
oscillatory pattern of the acoustic peaks.  Unfortunately, the sample variance of these estimators
is also higher given the imperfect correlation between the two fields.

On the other hand, the sample variance of the $EB$ estimator is reduced 
by the fact that $B$ modes are assumed absent in the unlensed fields \cite{HuOka01,HirSel02}.   
In practice the noise
of the actual estimator will depend on the $B$-modes contributed by foregrounds and other contaminants.
Like the $ET$ estimator, cross correlation of the large scale $E$ gradients with the small scale $B$-modes
should help reduce any bias arising from contaminants.
Since the experimental requirements for $EB$ are similar to that of $TE$, $TB$, and $EE$ but yield better prospects
for constraints, we focus on the $EB$, $ET$ and $TT$ estimators in the next section.

\section{Idealized Examples}
\label{sec:examples}

To illustrate the performance of the estimators, let us take the idealization that all lenses are
 NFW \cite{NavFreWhi97} dark matter halos and the observed CMB fields have no
 contaminants aside from white detector noise.   We will address more realistic cases
 in a future work \cite{ValDeDHu07}. The estimators should remain nearly unbiased
  since to pass through the reconstruction filters, a contaminant must be antisymmetric
 around the cluster center with an axis whose direction is correlated with the large scale gradients.
 In addition, contaminants will not appear in the 6 quadratic estimators, especially
 those involving cross correlation, in the same way.
    They can however contribute substantially to the noise and we
 will examine here the performance of the estimators as a function of detector
  noise as a proxy for
 other contaminants.

The NFW density profile is given by  \cite{NavFreWhi97}
\begin{equation}
\rho(r) \propto {1 \over r/r_s(1+r/r_s)^2} \,,
\end{equation}
where the normalization coefficient can be expressed in terms of
 the mass of the halo.  We define the mass to be that
enclosed at $r_{180}$ defined to be the radius which encloses the mass at an 
overdensity of 180 times the {\it mean} density
\begin{equation}
r_{180} = \left(  {3 M \over 4\pi} {1\over {180\Omega_m\rho_c} } \right)^{1/3}\,,
\end{equation}
where $\rho_c$ is the critical density today. 
Likewise, we define the concentration of the halo in terms of this radius
\begin{equation}
 c_{180}= {r_{180} \over r_s  }\,.
\end{equation}
The convergence profile for such a halo is \cite{Bar96}
\begin{equation}
\kappa(\theta) = {3 \over 4\pi} {H_0 D_L}{D_{LS} \over {D_S}} (1+z_L) {M  H_0 \over
\rho_c r_s^2 } {g(\theta /\theta_s) \over f({c_{180}})}\,,
\end{equation}
where the projected scale radius $\theta_s = r_s/D_L$. $D$ is the comoving
distance in a flat universe with subscripts ``L" denoting the distance to the lens, ``S" denoting
distance to the source, and ``LS" the distance between the lens and source.  A primary
advantage of the CMB is that the source is behind all clusters and at a well-determined
distance $D_S =14.3$Gpc.
The concentration factor is
\begin{equation}
f(c) = \ln (1+c) - {c \over 1+c}\,,
\end{equation}
and the functional form of the profile is \cite{Bar96}
\begin{eqnarray}
g(x)& =& { 1 \over x^2-1} \left[ 1- {2 \over \sqrt{x^2-1}} {\rm atan} \sqrt{ {x-1 \over x+1}}\right],  \quad (x>1) 
\nonumber\\
&=&  { 1 \over x^2-1} \left[ 1- {2 \over \sqrt{1-x^2}} {\rm atanh} \sqrt{ {1-x \over1+x}}\right], \quad (x<1)
\nonumber\\
&=&  { 1 \over 3}\,,  \quad (x=1)\,.
\end{eqnarray}
For illustrative purposes, we take 
\begin{equation}
M_{14} \equiv {M \over 10^{14} h^{-1} M_\odot} = 2 \,,
\end{equation}
$c_{180} = 3.2$
and $z_L=0.7$.  For these parameters $\theta_s = 0.94'$ and $\kappa(\theta_s)= 0.1$.
Note the low value of the concentration reflects the low $\sigma_8$ fiducial
cosmology and is determined by the scaling of \cite{Buletal01}.  In a higher $\sigma_8$ cosmology,
clusters at a fixed mass are more concentrated but correspondingly larger clusters are more abundant.
Other numbers are typical of SZ selected clusters  with upcoming surveys.

\begin{figure}[th]
\centerline{\psfig{file=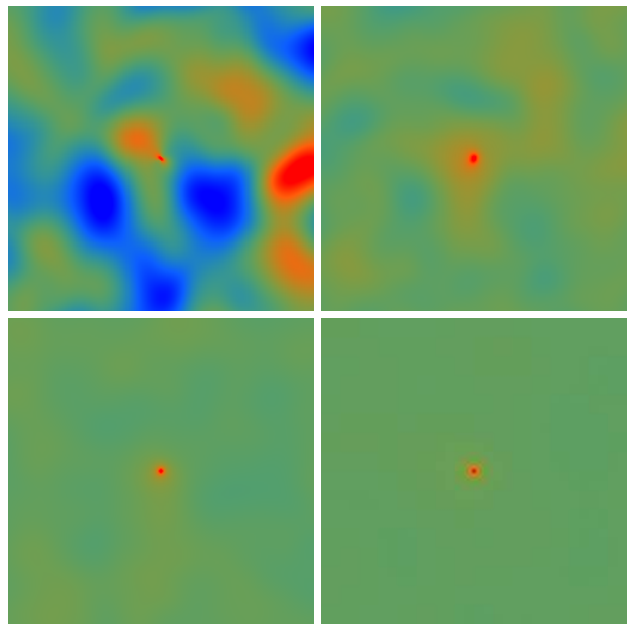, width=3.25in}}
  \caption{Temperature ($TT$) reconstruction in the absence of detector noise and other contaminants.  
  From left to right, top to bottom
  the reconstructions are stacked with 1, 10, $10^2$, $10^3$
   identical clusters of $M_{14}= M/10^{14}h^{-1}M_\odot=2$ and $z=0.7$.  Shown here is a $35' \times 35'$ region around the cluster.}
  \label{fig:recon}
\end{figure}

For the simulated reconstruction we consider a fiducial experiment with a $\theta_{\rm FWHM}=1'$
beam and varying levels of noise on the sky
\begin{eqnarray}
N_l^{TT} &= &  \Delta_T^2 B_l^{-2} \,, \nonumber\\
N_l^{EE} & =& N_l^{BB} = \Delta_P^2 B_l^{-2} \,.
\end{eqnarray}
where the beam factor is
\begin{equation}
B_l = 
 e^{-l^2\theta_{\rm FWHM}^2/16\ln 2}
 \end{equation}
 with $\theta_{\rm FWHM}$ in radians.  It exponentiates the white detector noise on the beam deconvolved sky.
 We further assume that $\Delta_P = \sqrt{2} \Delta_T$.
For this fiducial beam, a pixel scale of  $\theta_{\rm pix}=0.2'$ suffices.  The field is
chosen to be 
$512\times 512$ pixels $(100' \times 100')$. 

We lens realizations of
 $\tilde C_l^{TT}$, $\tilde C_l^{EE}$ according to the full non-linear prescription of
Eq.~(\ref{eqn:mapping}) with 6th order polynomial interpolation of the source image.
This procedure allows strong lensing effects to appear at the
center of the cluster.  To this
lensed image we add a realization of the noise power spectra above.

In  Fig.~\ref{fig:recon}, we show the reconstruction with $\Delta_T=0$ simulations 
of the 
$TT$-estimator for $N=1$, 10, $10^2$, and $10^3$ stacked clusters.   In this detector noise free limit,
there is a clear detection with only one cluster.  It appears as a concentrated spike on top of
a noisy but slowly varying background from chance correlations of the Gaussian random background field.
The reconstructed single image lacks the azimuthal symmetry of the lens but symmetry is
recovered by stacking images \cite{MatBarMenMos05}.  This asymmetry comes from  the fact that 
the deflections cannot be estimated in the direction orthogonal
to the gradient.  The estimator however is unbiased once many realizations of the
gradient are stacked.

\begin{figure}[th]
\centerline{\psfig{file=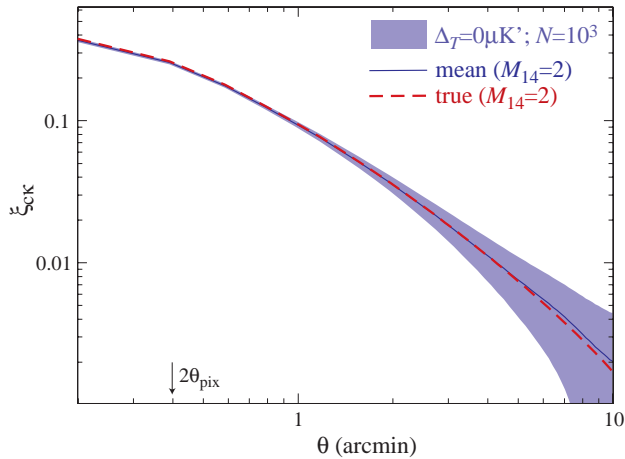, width=3.25in}}
  \caption{Recovered cluster-mass correlation function or stacked profile with 24000 detector noise
  free simulations.  Bands represent the rms fluctuation per 1000 clusters in the sample.  For $M_{14}=2$, the 
  reconstruction has no detectable bias at the $1-2\%$ level within a few arcmin.}
  \label{fig:unbiased}
\end{figure}

Stacking is equivalent to measuring the cluster-convergence correlation function
\begin{equation}
\xi_{c\kappa}(\theta)= \int{ d\phi \over 2\pi}\langle   \hat \kappa(\theta,\phi) \rangle\,,
\end{equation}
where the coordinates are centered at the location of the cluster such that $\theta$ is
the separation from the center and $\phi$ is the azimuthal angle around the cluster.
The averaging is over the clusters in the sample.  By defining the observable as
the cluster-convergence correlation function, the estimator naturally generalizes
to non-identical clusters and projected mass along the line of sight associated with
the cluster.  In practice, we evaluate the correlation function at discrete intervals
in pixel units by interpolation on the stacked image.  Neighboring bins out to several arcminutes
 are therefore
highly correlated as can be seen directly in Fig.~\ref{fig:recon}.  This correlation is in fact useful
for distinguishing the signal from noise.

In Fig.~\ref{fig:unbiased}, we show the recovered correlation function from $24000$ detector noise free
realizations along with the scatter per 1000 clusters with $M_{14}=2$.  We have smoothed both the lens and reconstruction with
a $1.5$ pixel FWHM beam after the fact for comparison.
 The estimator has no detectable bias at the $\sim 1-2\%$
level in the well-measured regime within a few scale radii.  At five times the fiducial mass or $M_{14}=10$, a low bias
of $\sim 8-9\%$ develops for the fiducial gradient cut of $l_G=2000$.  For these very rare clusters,
the signal is large enough that a more aggressive $l_G=1000$ yields a bias free reconstruction
with sufficient signal to noise.  Alternately, the bias can be calibrated out with simulations.

 \begin{figure}[th]
\centerline{\psfig{file=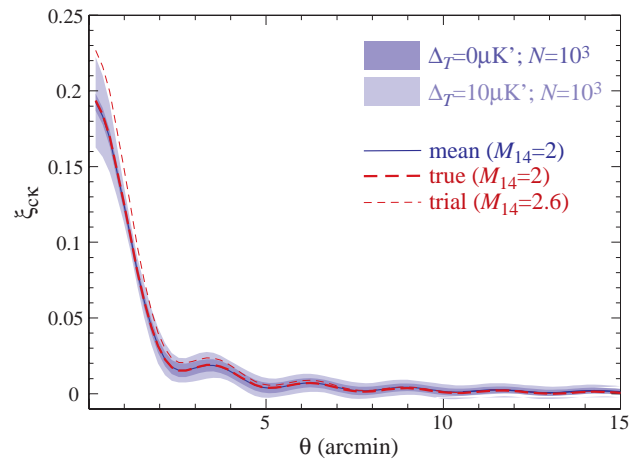, width=3.25in}}
  \caption{Beam filtered correlation function from 12000 clusters.  
  The reconstruction is low
  pass filtered at  $l<l_{\rm beam}(=8095)$ to remove contributions below the beam scale.   
  Shown for comparison are the filtered input lens of $M_{14}=2$ and a trial model of $M_{14}=2.6$ for
  comparison.  With a noise level of $10\mu$K$'$ the two cases are clearly distinguished.}
  \label{fig:filtered}
\end{figure}

For cases with finite noise, we explicitly remove effects below the beam scale for numerical
convenience.   We low pass filter both the lens and the reconstruction with a step function at
$l_{\rm beam} = \sqrt{8\ln 2} /\theta_{\rm FWHM}$.   This also has the desirable effect of 
making the reconstruction 
less sensitive to potential contamination near the cluster center.   The filtered correlation function is shown in
Fig.~\ref{fig:filtered} and remains unbiased and well-defined.   It is equivalent to measuring
the cross power spectrum at multipoles $l < l_{\rm beam}$.   Fig.~\ref{fig:filtered} also shows that with
 $\Delta_T = 10\mu$K$'$, $M_{14}=2$ and $2.6$ are significantly distinguished.

 To quantify this sensitivity, we compute the covariance matrix of the correlation function 
 from the simulations  
\begin{equation}
C(\theta,\theta') = \langle\hat  \xi_{c\kappa}(\theta)\hat \xi_{c\kappa}(\theta')\rangle 
-  \xi_{c\kappa}(\theta)   \xi_{c\kappa}(\theta')\,,
\end{equation}
where the discretization again is in bins of the pixel width.  
To estimate the significance with which two models can be separated, we evaluate
$\Delta\chi^2$ between a test model $\xi_{c\kappa}'(\theta)$ and the true model
$\xi_{c\kappa}(\theta)$
\begin{equation}
\Delta \chi^2 = \sum_{\theta,\theta'} [\xi_{c\kappa}'(\theta) - \xi_{c\kappa}(\theta)] {\bf C}^{-1}
[\xi_{c\kappa}'(\theta')-\xi_{c\kappa}(\theta')] \,,
\end{equation}
summed out to $\theta=5'$.  Beyond a few scale radii of the cluster
the correlation function is dominated by large scale structure.

\begin{figure}[ht]
\centerline{\psfig{file=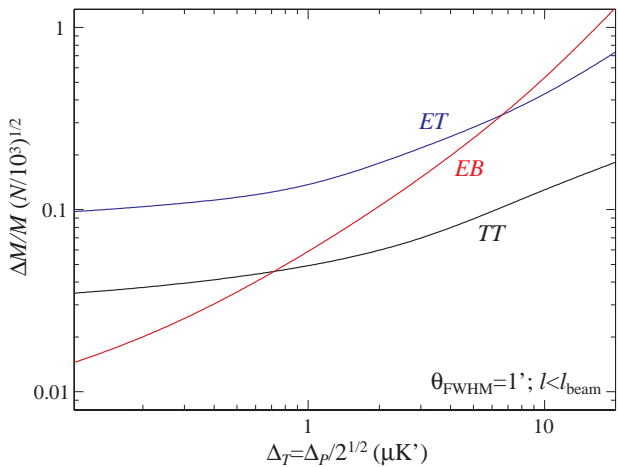, width=3.25in}}
  \caption{Idealized mass sensitivity for $\theta_{\rm FWHM}=1'$, 
  $\theta \le 5'$.   
  The fractional errors on the idealized NFW mass with 1000 clusters
  for the $TT$, $ET$ and $EB$ estimators.  Unrelated mass along the line of sight will add noise
  to the estimates as will any emission from the cluster.  
  No information from the reconstruction 
  is used below the beam scale $l \ge l_{\rm beam}$.}
  \label{fig:masscal}
\end{figure}

In Tab.~\ref{tab:sig}, we show the detection significance, the difference between the fiducial $M_{14}=2$ lens
and zero with the covariance matrix evaluated at zero signal.  We have scaled the
significance per cluster by $\sqrt{N}$ for $N=10^3$.  
At a noise level of $10\mu$K$'$,
the significance or signal to noise is $\sqrt{\Delta \chi^2} \sim 0.4$ per cluster.    Above this noise level, only $TT$
and $ET$ can provide significant detections.  Although $TT$ has 
roughly three times the signal-to-noise,
$ET$ can still provide a useful check against contamination and systematics for a first
detection.  Recall that the filter construction involves the oscillatory acoustic $TE$ correlation
and strongly rejects spurious signals.  At noise levels between  $1-3\mu$K$'$, the $EB$ estimator
has both the signal-to-noise and the ability to reject contaminants to make it competitive with $TT$.
Below $1\mu$K$'$, the $EB$ estimator dominates.   

In Fig.~\ref{fig:masscal} we plot the sensitivity to the cluster mass. Specifically we calculate
the fractional change in mass that would generate a $\Delta \chi^2=1$
with $10^3$ clusters.  The covariance is  evaluated at the true model $M_{14}=2$ so as to include
the sample variance of the unlensed gradients.  
The same features found for detection significance hold for mass sensitivity.  For example 
at $10\mu$K$'$,  $TT$, $ET$, and $EB$ provide 13\%, 43\%, 53\% mass estimates
whereas at $1\mu$K$'$ the numbers improve to 5\%, 14\%, 6\%.

Finally, these sensitivities depend on the beam scale only weakly near the fiducial
$\theta_{\rm FWHM}=1'$ and $\Delta_T =1-10\mu$K$'$ level.   For example, at $3\mu$K$'$,
improving the beam to $0.5'$ has negligible effect on the signal to noise.  Enlarging the
beam to $2'$, degrades the $TT$ and $ET$ estimators by $\sim 30-40\%$ in mass sensitivity. 
Since the $EB$ estimator is noise dominated at the $1'$ scale even without beam, it degrades
negligibly.

\begin{table}[ht]
\caption{\label{tab:sig}Detection significance $\sqrt{\Delta \chi^2}$
or S/N for $10^3$ clusters of $M_{14}=2$, $z=0.7$ and
$\theta_{\rm FWHM}=1'$.}
%\begin{tabular*}{3.25in}{@{}l*{15}{@{\extracolsep{0pt plus
%12pt}}l}}
%\br
\begin{tabular}{l|rrr}
Type & $3\mu$K$'$ & $10\mu$K$'$ & 30$\mu$K$'$\\
\hline
%\mr
$TT$ & 22.9 & 11.5 & 6.1 \\
$ET$ &8.0 & 3.5 &1.4 \\
$EB$ &8.6 & 2.4 & 0.6 \\
%\br
\end{tabular}
\end{table}

\section{Discussion}

We have developed mass estimators from 6 quadratic combinations of CMB temperature and 
polarization fields that retain their minimum variance characteristic for large scale structure
and provide nearly unbiased reconstructions for rare clusters where lensing effects are moderate
to strong.  The key difference from previous work \cite{Hu01b,HuOka01} is that the Wiener filtering
for unlensed CMB gradients is augmented with a sharp filter that removes any
gradients at $l_G>2000$ that are not part of the source fields from recombination.

This sharp filter should also help prevent false signals from cluster emission such as the 
thermal and kinetic Sunyaev-Zel'dovich effect, point source emission as well as
other foregrounds.  To even appear as excess noise, 
the
contaminant at $\sim 1'$ must have a component that is antisymmetric about the cluster center.
To bias the measurement, the direction of the asymmetry must be 
correlated with both the polarized and unpolarized contamination
 at $\sim 10'$  in the same way as the acoustic oscillations in the CMB.  Note that 
 projection effects from mass associated with the cluster should be considered part
 of the signal and can be calibrated by $N$-body simulations.
 True contaminants however can substantially increase the noise of the estimators.
An evaluation of the efficacy of the filter in the presence of real world contaminants
is beyond the scope of this paper but will be treated in 
future work \cite{ValDeDHu07}.

Here we have tested the estimators against idealized signal and noise: identical clusters lenses
with NFW dark matter profiles in the presence of beam filtered white noise.  
At a noise level of $10\mu$K$'$ and with a $1'$ beam, 
the $TT$ estimator can provide a $\sim 10\sigma$ detection
per $10^{3}$ clusters of $M \sim 2 \times 10^{14} h^{-1} M_\odot$ or equivalently a $\sim 10\%$
average mass measurement.   The $ET$ estimator, based on a separate measurement of $E$-polarization
in the acoustic peaks and $T$ measurements around the cluster, can provide an important
cross check on a first detection.   Since the estimator filters for the $TE$ correlation and
anticorrelation of the acoustic peaks it provides a strong discriminant against contamination
and systematics at the price of a factor of $\sim 3$ in signal-to-noise.  Ultimately with
experiments that produce foreground free maps of the polarization to $< 3\mu$K$'$, the
$EB$ estimator will return the best estimates, with the potential to ultimately 
provide $\sim 1\%$ measurements.

\bigskip 

{\it Acknowledgments} We thank Tom Crawford, Gil Holder,  Kendrick Smith, Scott Dodelson, 
 and Jeremy Tinker for useful conversations.
W.H. and S.D. were supported  by  the KICP under NSF PHY-0114422.
W.H. was additionally supported by 
U.S. Deptartment of Energy contract DE-FG02-90ER-40560 and the 
David and Lucile Packard Foundation. 
C.V. was supported by the U.S. Department of Energy and by NASA grant NAG5-10842.

\newpage

\bibliography{HuDeDVal06}

\end{document}